# The Case for Cross-Layer Optimizations in Storage: A Workflow-Optimized Storage System

S. Al-Kiswany, E. Vairavanathan, L. B. Costa, H. Yang, M. Ripeanu

## ABSTRACT

This paper proposes using file system custom metadata as a bidirectional communication channel between applications and the storage system. This channel can be used to pass hints that enable cross-layer optimizations, an option hindered today by the ossified file-system interface. We study this approach in context of storage system support for large-scale workflow execution systems: Our workflow optimized storage system (WOSS), exploits application hints to provide per-file optimized operations, and exposes data location to enable location-aware scheduling.

This paper argues that an incremental adoption path for adopting cross-layer optimizations in storage systems exists, presents the system architecture for a workflow-optimized storage system and its integration with a workflow runtime engine, and evaluates the proposed approach using synthetic as well as real applications workloads.

## 1. INTRODUCTION

Custom metadata features (a.k.a., 'tagging') have seen increased adoption in systems that support the storage, management, and analysis of 'big-data' [1, 2, 3, 4]. However, the benefits expected are all essentially realized at the application level either by using metadata to present richer or differently organized information to users (e.g., better search and navigability [5, 6]) or by implicitly communicating among applications that use the same data items (e.g., support for provenance [7]).

**Our thesis** is that, besides the above uses, custom metadata can be used as a *bidirectional communication channel* between applications and the storage system and thus *become the key enabler for cross-layer optimizations* that, today, are hindered by an ossified file-system interface.

This communication channel is *bidirectional* as the cross-layer optimizations enabled are based on information passed in both directions across the storage system interface (i.e., application to storage and storage to application). Possible cross-layer optimizations (surveyed in detail in §5) include:

- (top-down) Applications can use metadata to provide hints to the storage system about their future behavior, such as: per-file access patterns, ideal data placement (e.g. co-usage), predicted file lifetime (i.e., temporary files vs. persistent results), access locality in distributed setting, desired file replication level, or desired QoS. These hints can be used to optimize the storage layer.
- (bottom-up) The storage system can use metadata as a mechanism to expose key attributes of the data items stored. For example, a distributed storage system can provide information about data location, thus enabling location-aware scheduling.

The approach we propose has three interrelated advantages: it uses an application-agnostic mechanism, it is incremental, and it offers a low cost for experimentation. First, the communication mechanism we propose: simply annotating files with arbitrary *<key, value>* pairs, is application-agnostic as there are no application-specific provisions for cross-layer information passing. Second, our approach enables evolving applications and storage-systems independently while maintaining the current interface (e.g., POSIX), and offers an incremental transition path for legacy applications and storage-systems: A legacy application will still work without changes (yet will not see performance gains) when deployed over a new storage system that supports cross-layer optimizations. Similarly a legacy storage will still support applications that attempt to convey optimization hints, yet will not offer performance benefits. As storage and applications incrementally add support for passing and reacting to optimization hints, the overall system will see increasing gains. Finally, exposing information between different system layers implies tradeoffs between performance and transparency. To date, these tradeoffs have been scarcely explored. We posit that a flexible encoding (key/value pairs) as the information passing mechanism offers the flexibility to enable low-cost experimentation within this tradeoff space.

The approach we propose falls in the category of 'guided mechanisms' (i.e., solutions for applications to influence data placement, layout, and lifecycle), the focus of other projects as well. In effect, the wide range (and incompatibility!) of past such solutions proposed in the storage space in the past two decades (and incorporated to some degree by production systems - pNFS, PVFS, GPFS, Lustre, and other research projects [8, 9, 10, 11, 12, 13]), only highlights that *adopting an unifying abstraction is an area of high potential impact*. The novelty of this paper comes from the "*elegant simplicity*" of the solution we propose. Unlike past work, we maintain the existing API (predominantly POSIX compatible), and, within this API, we propose using the existing extended file attributes as a flexible, application-agnostic mechanism to pass information across the application/storage divide.

This work demonstrates that significant improvements are possible, without abandoning POSIX and that it is feasible to build a POSIX compliant storage system optimized for each application (or application mix) even if the application exhibits a heterogeneous data access pattern.

***We demonstrate our approach*** by building a POSIX-compatible storage system to efficiently support one application domain: scientific workflows. We chose this domain as this community has to support a large set of legacy applications (developed using the POSIX API). The storage system aggregates the resources of the computing nodes allocated to a batch application (*e.g.*, disks, SSDs, and memory) and offers a shared file-system abstraction with two key features. First, it is able to efficiently support the data access patterns generated by workflows through file-level optimizations. To this end, the storage system takes hints that offer information about the expected access pattern on a specific data item or collection of items and guides the data layout (e.g., file and block placement, file co-placement). Second, the storage system uses custom metadata to expose data location information so that the workflow runtime engine can make location-aware scheduling decisions. These two features are key to efficiently support workflow applications as their generated data access patterns are irregular and application-dependent.

*The key contributions of this work are:*
- We propose a new approach that uses custom metadata to enable cross-layer optimizations between applications and the storage system. Further, we argue that an incremental adoption path exists for adopting this approach. This suggests an evolution path for co-designing POSIX-compatible file-systems together with the middleware ecosystem they coexist within such that performance efficiencies are not lost and flexibility is preserved, a key concern when aiming to support legacy applications.

- To demonstrate the viability of this approach, we present the design of a workflow-optimized storage system (WOSS) based on this approach. This design provides generic storage system building blocks that can be adopted to support a wider range of cross-layer optimizations. Based on these building blocks, our design supports data access patterns frequently generated by workflows by enabling the workflow runtime engine to pass per-file/collection access hints and the storage to expose data location and thus enable location-aware task scheduling. Importantly, we argue that it is possible to achieve our goals without changing the application code or tasking the application developer to annotate their code to reveal the data usage patterns.

- We offer an open-source implementation of the system and we have integrated it with two workflow runtime engines (*pyFlow*, developed by ourselves, and *Swift* [14]). On the storage side, we have started from an existing object-based storage system (MosaStore http://mosastore.net) and added the ability to offer and react to hints. On the workflow runtime side, we have added data-location aware scheduling.

- We demonstrate, using synthetic benchmarks as well as three real-world workflows that this design brings sizeable performance gains. On a commodity cluster, the synthetic benchmarks reveal that, compared to a traditionally designed distributed storage system that uses the same hardware resources, WOSS achieves from 30% to up to 2x higher performance depending on the access pattern. Further, compared to a NFS server deployed on a well provisioned server-class machine (with multiple disks, and large memory), WOSS achieves up to 10x performance gains. (NFS only provided competitive performance under cache friendly workloads) Further, under real applications, WOSS enabled an 20-30% application-level performance gain, and 30-50% gain compared to NFS. Finally, our evaluation a Blue Gene/P machine shows that WOSS can scale to support larger workloads and enables sizable gains compared to the deployed backend storage (GPFS).

*Relationship to our own past work.* We have originally presented the idea of using custom-metadata to enable cross-layer optimizations in a storage system in a 'hot-topic' paper at HPDC'08 [15]. Additionally, we have used [16] synthetic benchmarks and small-scale experiments to convince ourselves that per-file optimizations have the potential to bring benefits in practice. For these experiments, however, we have not built a system prototype, but just 'hacked' MosaStore. This is the first time we report on a complete system design, build a prototype with complete functionality, integrate it with workflow runtime engines, and evaluate it at scale and with real applications.

*Organization of this paper.* The final section of this paper includes a detailed design discussion and design guidelines, discusses the limitations of this approach, and elaborates on the argument that custom metadata can benefit *generic* storage systems by enabling cross-layer optimizations (§5). Before that, we present the context (§2), the design (§3) and evaluation (§4) of a first storage system we designed in this style: *the workflow-optimized storage system (WOSS)*.

## 2. BACKGROUND AND RELATED WORK
This section starts by briefly setting up the context: the application domain and the usage scenario we target. It then continues with a summary of data access patterns of workflow applications and a survey of related work on alleviating the storage bottleneck.

*The application domain: workflow applications.* Meta-applications that assemble complex processing workflows using existing applications as their building blocks are increasingly popular in the science domain. While there are multiple ways to support the execution of these workflows, in the science area — where a significant legacy codebase exists — one approach has gained widespread adoption: a *many-task approach* [20] in which meta-applications are assembled as workflows of independent, processes that communicate through intermediary files.

There are three main advantages that make most workflow runtime engines adopt this approach and use a shared file-system to store the intermediary files: simplicity, direct support for legacy applications, and support for fault-tolerance. First, a shared file-system approach simplifies workflow development, deployment, and debugging: essentially workflows can be developed on a workstation then deployed on a large machine without changing the environment. Moreover, a shared file-system system simplifies workflow debugging as intermediate computation state can be easily inspected at runtime and, if needed, collected for debugging or performance profiling. Second, a shared file-system will support the legacy applications that form the individual workflow stages as these generally use the POSIX API. Finally, compared to approaches based on message passing, communicating between workflow stages through a storage system that offers persistency makes support for fault-tolerance much simpler: a failed execution step can simply be restarted on a different compute node as long as all its input data is available in the shared file-system.

Although these are important advantages, the main drawback of this approach is low performance: the file-system abstraction constrains the ability to harness performance-oriented optimizations that can only be provided if information is shared between system layers (Figure 1). More specifically, a traditional file system cannot use the information available at the level of the workflow execution engine (e.g., to guide the data placement) Similarly, as traditional file-systems do not expose data-location info, the workflow runtime engine cannot exploit opportunities for collocating data and computation.

**Usage scenario: batch applications**. Since, on large machines, the back-end file-system becomes a bottleneck when supporting I/O intensive workflows [21, 22], today's common way to run them is to harness some of the resources allocated by the batch-scheduler to the application and assemble a scratch shared file-system that will store the intermediary files used to communicate among workflow tasks. This usage scenario is similar to the one explored by BAD-FS [22]: the file system acts as a scratch space and offers persistence only for the duration of the application; input data and results are staged-in/out.

It is this batch-oriented scenario, described in more detail in Figure 1 and its legend, that we assume for the rest of the paper. We note that the shard file-system offered, facilitates integration and has become popular in other scenarios as well (e.g., checkpointing [23], stage-in/out [24], analytics [25], in-memory analysis of intermediate results, visualization).

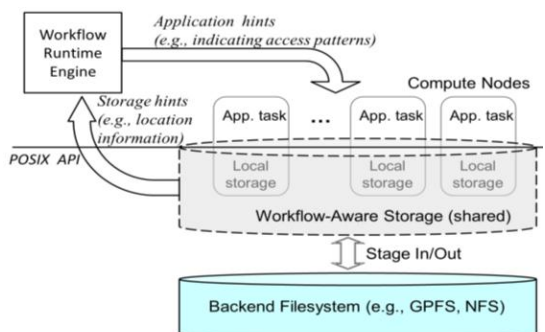

**Figure 1**. **Usage scenario and high-level architecture.** The workflow optimized storage system (WOSS) aggregates the storage space of the compute nodes and is used as an intermediate file-system. Input/output data is staged in/out from the backend storage. The workflow scheduler queries WOSS for data location to preform location-aware scheduling. The scheduler submits tasks to individual compute nodes and includes hints on the data usage patterns.

### 2.1. Comon Workflow Data Access Patterns

Several studies explore the data access patterns of scientific workflows: [26, 27, 28, 29, 30]. To make this paper self-contained, this subsection briefly presents the common patterns identified by these studies, the opportunities for optimizations they generate, and the support required from a storage system. Due to constraints, we detail below only a few of the patterns, while Table 1 is summarizes all:

- *Pipeline*: A set of compute tasks are chained in a sequence such that the output of one task is the input of the next task in the chain. An optimized system will store an intermediate output file on a storage node on the same machine as the one that executes the task producing it (if space is available) to increase access locality and efficiently use local caches. Ideally, the location of the data is exposed to the workflow scheduler so that the task that consumes this data is scheduled on the same node.
- *Broadcast*: A single file is used by multiple tasks. . An optimized storage system can create enough replicas of the shared file to eliminate the possibility that the node(s) storing the file become overloaded..
- *Reduce*: A single compute task uses as inputs files

**Table 1. Common workflow patterns.** Circles represent computations. *An outgoing arrow indicates that data is produced (to a temporary file) while an incoming arrow indicates that data is consumed (from a temporary file).* There may be multiple inputs and outputs via multiple files. (Notation similar to that used by Wozniak *et al.* [26]). Arrows are labeled with extended attribute API calls used to pass hints to enable the optimizations. (The corresponding hints are presented in detail in Table 3)

| Pattern | Pattern Details | Optimizations / Hint |
|---|---|---|
| Pipeline | Input → set(dp_local) → get(file, location) → set(dp_local) → get(file, location) → Output | ▪ Node-local data placement (if possible). ▪ Caching. ▪ Data location-aware scheduling. |
| Broadcast | Input → set(replicate, 4) → get(file, location) → Output | ▪ Optimized replication taking into account the data size, the fan-out, and the topology of the interconnect. |
| Reduce | Input → set(dp_collocate) → get(file, location) → Output | ▪ Reduce-aware data placement: co-placement of all output files on a single node; ▪ Data location-aware scheduling |
| Scatter | Input file → set(block-size, size) set(dp_striped) → get(file, location) → Output | ▪ Application-informed block size for the file. ▪ Application-aware block placement; ▪ Data-location application scheduling |
| Gather | set(block-size, size) set(dp_chunk_local) → Output file | ▪ Application-informed block size for the file. ▪ Application-aware block placement. |
| Reuse | Input file → Tasks on the same node → Output | ▪ Application-informed replication. ▪ Application-informed caching. |
| Distribute | Input → Output files | ▪ Application informed file and chunk placement. ▪ Application informed replication. |

produced by multiple computations. Examples include a task that checks the results of previous tasks for a convergence criterion, or a task that calculates summary statistics from the output of many tasks. An optimized storage system can intelligently place all these input files on one node and expose their location, thus creating an opportunity for scheduling the reduce task on that node to increase data access locality.

### 2.2. Past Work on Alleviating the Storage Bottleneck
A number of alternative approaches have been proposed to alleviate the storage bottleneck for workflow applications. Taken in isolation, these efforts do not fully address the problem we face as they are either too specific to a class of applications; or enable optimizations system-wide and throughout the application runtime, thus inefficiently supporting applications that have different usage patterns for different files. Our solution integrates lessons from this past work and demonstrates that it is feasible to provide runtime storage optimizations per data-item.

*Application-optimized storage systems.* Building storage systems geared for a particular class of I/O operations or for a specific access pattern is not uncommon. For example, the Google file system [31] optimizes for large datasets and append access, HDFS [32] and GPFS-SNC [33] optimize for immutable data sets, location-aware scheduling, and rack-aware fault tolerance; the log-structured file system [34] optimizes for write-intensive workloads, arguing that most reads are served by ever increasing memory caches; finally BAD-FS [22] optimizes for batch job submission patterns. These storage systems and the many others that take a similar approach are optimized for one specific access pattern and consequently are inefficient when different data objects have different patterns, like in the case of workflows.

*Custom metadata in storage systems.* A number of systems propose mechanisms to efficiently support custom metadata operations including Metafs [35], Haystack [36], The Linking File System (LiFS) [4] and faceted search [5]. These systems extend the traditional file system interface with a metadata interface that allows applications to create arbitrary metadata. These efforts provide applications the functionality of annotating files with arbitrary <key, value> pairs and/or to express relationships among files. Similarly, Graffiti [37] is a middleware that allows tagging and sharing of tags between desktop file systems. As other systems which aim to provide a metadata interface, it supports tags and links between files, but focuses on sharing-related issues. These solutions essentially use metadata to communicate between applications. They focus on providing better data search, navigability, and organization at the application layer.

*Dealing with a constraining storage system interface.* Two solutions are generally adopted to pass hints from applications to the storage system: either giving up the POSIX interface for a wider API, or, alternatively, extending POSIX API with an orthogonal additional ad-hoc interface for hint passing. Most storage systems that operate in the HPC space (pNFS, PVFS, GPFS, Lustre) fall in the latter category and add ad-hoc hint passing interfaces; while most Internet services/cloud storage systems fall (e.g., HDFS) fall in the former category. In terms of exposing data location the situation is similar: HDFS and other non-POSIX systems do expose data location to applications while most parallel large-scale file systems (e.g., pNFS, PVFS, GPFS) do not expose it (even though this information may be available at the client module level) [38]. Further we note that these systems cannot support cross layer optimizations as their design does not support per-file optimizations, does not have mechanisms to enable/disable optimizations based on application triggers, nor allows extending the system with new optimizations.

*Storage system optimizations using application provided hints.* A number of projects propose exploiting application information to optimize the storage system operations. Mesnier et. al. [39] propose changes to the storage system block API to classify storage blocks into different classes (metadata, journal, small/large file), allowing the storage system to apply per class QoS polices. Collaborative caching [40] proposes changing the storage system API to facilitate passing hints from clients to server to inform the server cache mechanism. Finally, Patterson et. al. [41] propose an encoding approach to list the blocks an application accesses, and to use the IO control interface (ioctl) to pass this list to the storage system which uses it to optimize caching and prefetching. For example eHiTS [42] and GreenStor [43] storage systems propose building energy optimized distributed storage system that use application hints to facilitate turning off or idling storage devices holding data blocks that will not be accessed in the near future.

BitDew [9], a programming framework for desktop grids, enables users to provide hints to the underlying data management system. The five supported hints are: replication level, resilience to faults, transfer protocol, data lifetime, and data affinity (used to group files together).

In this same vein, UrsaMinor [44], an object-based storage system, allows the system admin or the application, through a special API, to configure the storage system operations for its data objects. For each object, the system facilitates configuring the reliability mechanism (replication or erasure coding), and fault and timing model. Through this specialization the system better meets application requirements in terms of throughput and reliability.

Finally, the XAM [45] standard defines an extended API and access method for storage systems serving mostly immutable data (e.g. backup systems, email servers). It allows the programmer to better describe the data through extended metadata interface. Further it allows the programmer to inform the storage system of how long to retain the data through special metadata fields.

Table 2 compares this work to the related production and research projects. These efforts differ from our proposed approach in three main ways: First, most of them target a specific optimization and they do not build an extensible storage system that can be extended with new optimizations. Second, they propose uni-directional hint passing from application to storage Third, and most importantly, they either propose changes to the standard APIs to pass hints, or use current API (ioctl) in a non-portable, and non-standard way. This hinders the approaches adoptability and portability. Finally, no other project (Table 2) provides a bidirectional communication mechanism, proposes an extensible storage system design, while using a standard API. We note that all the optimizations proposed in the survived projects can use our cross layer communication mechanism to pass hints, and can be implemented using our system architecture.

**Table 2 Survey of related projects.** The table compares WOSS with current approaches on number of axes

| Projects | Domain specific / general | Production / research project | API | Bidirectional | Extensible |
|---|---|---|---|---|---|
| HDFS [32] | Domain | Production | New API | N | N |
| PVFS [46], GPFS [47] | Domain | Production | POSIX | N | N |
| GreenStor [43], TIP [48], | General | Research | New API | N | N |

| eHiTS [42] | | | | | |
|---|---|---|---|---|---|
| Mesnier et. al. [39] | General | Research | Modify Disk API | N | N |
| Collaborative caching [40] | General | Research | Non-portable POSIX | N | N |
| XAM [45] | General | Specification | New API | N | Y |
| BitDew [9] | Domain | Research | New API | N | N |
| WOSS | General | Research | POSIX compliant | Y | Y |

## 3. SYSTEM DESIGN

This section discusses the workflow optimized storage system (WOSS) design requirements, presents the system design, the prototype implementation, and the WOSS integration effort with *pyFlow* and *Swift* workflow runtimes.

### 3.1. Design Requirements
To efficiently support the usage scenario targeted and the access patterns generated by workflow applications (§2.0), WOSS needs to support the following requirements:

- *Extensibility*. The storage system architecture should be modular and extensible. For instance, it should be easy for a developer to define a new data placement policy that associated with a new custom attribute.
- *Fine-grain (e.g., per-file or collection of files) runtime configurability*: The storage system should provide per-file configuration at run time to support high-configurability for diverse applications access patterns. Further, the system should support defining a group of files and supporting per-group optimizations (e.g. collocation).
- *Deployable as an intermediate, temporary storage* that aggregates (some of) the resources allocated to the batch application. This will not only avoid potential backend storage performance and scalability bottlenecks, but will also enable location-aware scheduling as computation can be collocated with data. The storage system should be easy to deploy during the application's start-up. Further, ideally it should be transparently interposed between the application and the backend storage for automatic data pre-fetching or storing persistent data or results.
- *System-level configurability*: The storage system should provide system-wide configuration knobs to support configurability for diverse applications. The system should be tunable for a specific application workload and deployment. This includes ability to control local resource usage, in addition to controlling application-level storage system semantics, such as data consistency and reliability.
- *POSIX compatible,* to facilitate access to the intermediate storage space, without requiring changes to applications.
- *Support for chunking*. To support large files that do not fit in a single machine storage space, and to enable optimizations for scatter and gather patterns, the storage system should support dividing and storing a single file into multiple chunks.

### 3.2. Storage System Design
Our prototype is based on a traditional object-based distributed storage system architecture, with three main components (Figure 2): a centralized metadata manager, the storage nodes, and the client's system access interface (SAI) which provides the client-side POSIX file system interface. Each file is divided into fixed-size blocks that are stored on the storage nodes. The metadata manager maintains a block-map for each file. Table 3 lists the optimizations the current prototype implements and their associated hints/tags (POSIX extended attributes).

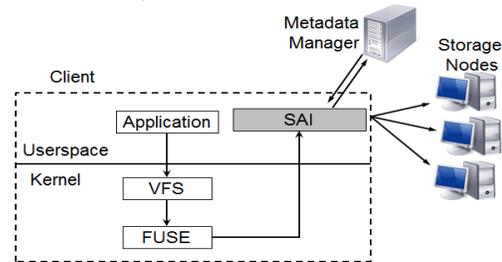

**Figure 2.The main components of a distributed storage system** (also used by WOSS): the metadata manager, the storage nodes, and the client module (detailed here, implemented on top of FUSE).

**Design for extensibility.** The distributed nature of the storage system makes providing hint-triggered optimizations challenging: while the hints (i.e., files' extended attributes) are maintained by the manager, the functionality corresponding to the various optimizations can reside at the manager (e.g., for data placement), client SAI (e.g., for caching), or storage nodes (e.g., for replication). Additionally we aim for a flexible design that supports exploration and facilitates adding new custom metadata and their corresponding functionality.

To this end, three main design decisions enable extensibility:
- *A generic hint propagation approach* that extends every message/request with optimization hints (i.e., it enables tagging communication messages) to enable propagating hints between the components (manager, storage node, and SAI). These per-message hints enable end-to-end information passing and optimized handling of every message/request across components.

    In our design, file-related operations are always initiated by the client SAI (e.g. space allocation, data replication request, or checking file attributes,). The first time an application opens a file or gets the file attributes, the SAI queries the metadata manager and caches the file's *extended* attributes (that carry the application hints). The SAI tags all subsequent internal inter-component communication related to that file (e.g., a space allocation, a request to store a data block) with the file's extended attributes and the callbacks that may be deployed at each component are triggered by these tags to implement the hint-directed optimizations.

- *Extensible storage system components design*. All storage system components follow a 'dispatcher' design pattern (Figure 3): all received requests are processed by the dispatcher and based on the requested operation and the associated hints (i.e., tags) the request/message maybe forwarded to the specific optimization module associated with the hint type, or processed using a default implementation.

    To extend the system with a new optimization for a specific operation (e.g., space allocation, replication, read, write …etc), the developer needs to decide the application hint (key-value pair) that will trigger the optimization, and implement the callback function the dispatcher will call when an operation on file with the associated hint is issued. Every optimization module can access the storage

component's internal information including reading the manager metadata or system status (e.g. storage nodes status) through a well-defined API, or, accessing the blocks stored at the storage nodes.

- *Passing hints bottom-up: an extensible information retrieval* design. To communicate a storage hint to the application the metadata manager provides an extensible information retrieval module (the GetAttrib module in Figure 3). This module is integrated with the dispatcher described in previous point as it is only triggered by the client POSIX 'get extended attribute' operation. Similar to other optimizations, to extend the system to expose specific internal state information the developer needs to decide the application hint/tag (key-value pair) that will trigger the optimization. The module, as all other optimization modules, has access to the manager metadata and system status information, and is able to extract and return to the client any internal information.

### 3.3. Prototype Implementation Details

We based our prototype implementation on MosaStore (http://mosastore.net) an existing distributed storage system. Our prototype changes MosaStore design and implementation to follow the extensible storage system design as described above. Similar to MosaStore, the WOSS SAI uses FUSE [49] kernel module to provide the POSIX file system interface.

We highlight a number of implementation details:

- *Replication operations* are carried by the storage nodes. Their design adopts a similar dispatcher architecture to enable multiple replication policies. In the current implementation, the application can select the replication policy and the number of replicas. The current implementation implements two replication policies: eager parallel replication (to replicate hot spot files as used in the broadcast pattern) and lazy chained replication (to achieve data reliability without increasing system overhead).
- *Exposing data location*: To expose files location our system defines a reserved extended attribute that has values for every file in the system ("location"). An application (in our case the workflow runtime) can 'get' the "location" extended attribute to obtain the set of storage nodes holding the file.

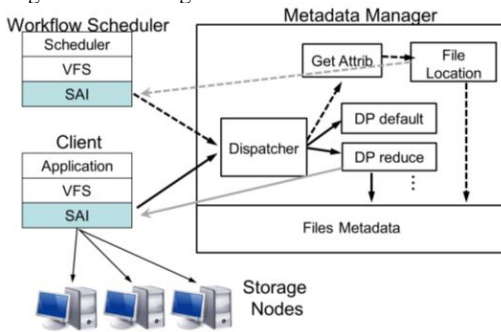

**Figure 3. WOSS metadata manager design**. For clarity, the figure shows WOSS integration with a workflow runtime engine and details WOSS metadata manager. The figure shows: (*i*) in solid lines, the path followed by a client chunk allocation request: the request is processed by a pattern-specific data placement 'DP' module based on the corresponding file tags/hints, (*ii*) the data path as data is produced by the clients (the solid lines going to storage nodes), and, (*iii*) the path of a request to retrieve file location information (dashed lines).

**Prototype limitations**. The prototype has three main limitations (All these limitations are the result of implementation decisions that enabled faster development and can be easily addressed with more resources). First, the prototype uses FUSE kernel module. While FUSE simplifies the SAI development it adds overhead to every file system call. Second, the data placement tags are only effective at file creation, changing the data placement tag for existing files will not change the file layout. Finally, our prototype uses a centralized metadata manager. While this introduces a potential bottleneck at scale, our experience is that the bottleneck that limited the overall system performance lied with the workflow runtime engine.

**Table 3. Implemented metadata attributes** (hints) and the corresponding optimizations

| Patterns and associated hints | Description |
|---|---|
| Pipeline pattern<br>set ("DP", "local") | Indicates preference to allocate the file blocks on the local storage node |
| Reduce pattern / collocation<br>set("DP", "collocation \|" <*group_name*>) | Preference to allocate the blocks for all files within the same <*group_name*> on the storage node. |
| Scatter pattern<br>set ("DP", "scatter \| <*scatterSize*>") | Place every group of contigues <*scatterSize*> chunks on a storage node in a round robin fashion |
| Broadcast pattern/replication<br>set("Replication",<*repNum*>) | Replicate the blocks of the file <*repNum*> times. |
| Replication semantics<br>Set("RepSmntc", "Optimisitc/Pessimestic") | Indicates which replication semantic to be used for the file: *optimistic*, return to application after creating the first replica, *pessimistic*, return to the application only after a chunk is well replicated. |
| Location<br>get ("location", null) | Retrieves the location information of the specific file. |
| Manage per file cache size<br>set("CacheSize",<size>) | Suggest a cache size per file (e.g. small cache size for small files or for read once files) |

### 3.4. Integration with a Workflow Runtime System

To demonstrate the end-to-end benefits of our approach, we integrated the WOSS prototype with *Swift*, a popular language and workflow runtime system [14] and with *pyFlow*, a similar, yet much simpler, system we have developed ourselves. (We stress that our integration does not require any modification to the application tasks). In particular we applied two modifications:

- *Adding location-aware scheduling*. The current implementations did not provide location-aware scheduling. We modified the schedulers to first query the metadata manager for location information, then attempt to schedule the task on the node holding the file. We note that our scheduling heuristics are relatively naïve, we estimate that further performance gains can be extracted with better heuristics; thus, our experiments provide a lower bound on the achievable performance gains.
- *Passing hints to indicate the data access patterns*. Information on the data access patterns is crucial to enable the ability of the storage system to optimize. Our experiments assume that the workflow runtime engine performs the task of determining the data access pattern as we see it as the most direct approach to obtain this information: The reason is that the runtime engine has access to the workflow definition, maintains the data dependency graph, and uses them to schedule computations. Thus, it already has the information to infer the usage patterns; the lifetime of each file involved, and can make computation placement decisions as well. Changing the workflow runtime implementation, however, to automatically extract this information is a significant development task (and not directly connected with the thesis we put forward here). Thus, we take a simpler approach: we inspect the workflow definitions for the applications we use in our evaluation and explicitly add the instructions to indicate the data access hints.

*Integration implementation limitations*: As one of our experiment highlights, our approach to integrate location-aware scheduling with *Swift* adds a significant overhead. This, for some scenarios at scale on BG/P, eliminates the performance gains brought by our optimizations. The problem here is that, to limit the changes we make in the *Swift* code, we implement every set-tag or get-location operation as a *Swift* task which, in turn, needs to be scheduled and launched in a computing node to call the corresponding POSIX command. With more time we can integrate this with *Swift*'s language and its Java-based implementation. The corresponding overhead is evaluated in §4.5 with *pyFlow*.

# 4. EVALUATION

We use a set of synthetic benchmarks and three real applications to evaluate the performance benefits of the proposed approach. To this end, we compare the proposed workflow-optimized storage system (labeled **WOSS** in the plots) with two baselines. First, as an intermediate storage scenario, we use MosaStore without any cross-layer optimizations (we label these experiments **DSS** – from distributed storage system to highlight that this is the performance we expect form a traditional object-based distributed storage system design). Since this setup is similar in terms of architecture and design to other cluster storage systems, such as Luster and PVFS [46], this comparison gives a rough estimate of the potential performance gains our technique can enabled. Second, we use a typical backend persistent storage system deployment (e.g., GPFS or NFS) available on clusters and supercomputers as another baseline. The reason for this additional baseline is to estimate the gains brought by the intermediate storage scenario and, additionally, to show that DSS is configured for good performance. Finally, where possible, we use a third baseline, the node-local storage, to expose the optimal performance achievable on the hardware setup.

To demonstrate that our approach and implementation are application-, workflow engine-, and platform-agnostic and that they bring performance improvements in multiple setups the synthetic benchmarks are implemented solely using shell scripts and *ssh* (secure shell) while the real applications use two workflow execution engines: *pyFlow* or *Swift*. These schedule the workflow tasks allocating the ready tasks to idle nodes according to the location information exposed by the storage system. Similarly, the shell scripts also query the storage system before launching a script on a specific machine. Finally, we evaluate using multiple platforms including a 20 nodes cluster and a BG/P machine

## Testbeds

We run most of our experiments on our lab cluster with 20 machines. Each machine has Intel Xeon E5345 4-core, 2.33-GHz CPU, 4-GB RAM, 1-Gbps NIC, and a RAID-1 on two 300-GB 7200-rpm SATA disks. The system has an additional NFS server as a backend storage solution that runs on a better provisioned machine with an Intel Xeon E5345 8-core, 2.33-GHz CPU, 8-GB RAM, 1-Gbps NIC, and a 6 SATA disks in a RAID 5 configuration. The NFS provides backend storage to the applications.

For two other larger scale experiments we use either 50 nodes on Grid5000 or one rack of an IBM BlueGene/P machine (850 MHz quad-core processors and 2GB RAM per node). The BG/P uses GPFS [47] as a backend storage system with 24 I/O servers (each with 20Gbps network connectivity). The computing nodes have no hard disks and mount a RAM disk. Details of the architecture can be found in [50].

When evaluating the DSS or WOSS systems, one node runs the metadata manager and the coordination scripts and the other nodes run the storage nodes (deployed over the local spinning disk or RAM disk), the client SAI, and the application executable.

### 4.1. Synthentic Benchmarks

Synthetic benchmarks provide relatively simple scenarios that highlight the potential impact of cross-layer optimizations on an intermediate storage scenario for each of the patterns described (Figure 4).

*Staging-in/out:* Current workflow systems generally use an intermediate storage scenario: they stage-in input data from a backend store (e.g., GPFS) to the intermediate shared storage space, process the data in this shared space, and then stage-out the results to persist them on the backend system. Overall, WOSS and DSS perform faster than NFS for the staging time and this section, although does not target evaluating staging, reports stage-in/out for the actual benchmark separately from the workflow time. *Note that adding the staging to the benchmark is conservative: these patterns often appear in the middle of a workflow application and, in those scenarios, staging would not affect them.*

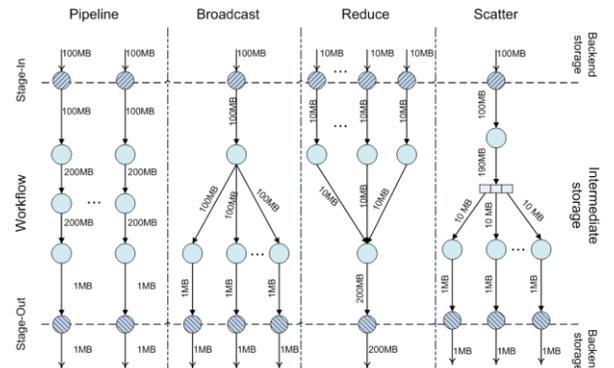

**Figure 4. Pipeline, broadcast, reduce, and scatter synthetic benchmarks.** Nodes represent workflow stages (or stage-in/out operations) and arrows represent data transfers through files. Labels on the arrows represent file sizes. Horizontal dashed lines represent the crossing boundary between backend and intermediate storage (e.g., stage-in reads a file from the backend and writes to the intermediate storage).

*Pipeline benchmark.* Each pipeline stages-in a common input file from the shared backend (i.e., the NFS server), goes through three processing stages using the intermediate storage and, then, the final output is staged out back to backend. The script tags the output files produced by a pipeline stage with a *'local'* tag to inform the storage to attempt storing the output files of a task on the node where the task runs. The storage exposes the file location so that the benchmark script can launch the next stage of the pipeline on the machine that already stores the file.

*Broadcast benchmark.* A single file is staged from NFS to intermediate storage. Then, a workflow stage produces a file in intermediate storage, which is consumed by 19 processes running in parallel, one per machine. When this file is created, the storage system creates eagerly (i.e., while each block is written) the number of replicas as specified by the replication tag. When the nodes process the input file, they randomly select a replica to read from, (giving preference to local blocks if available) avoiding a scenario where a storage node becomes a bottleneck. Each task produces, independently, an output file, and finally, these output files are staged-out to the backend storage.

*Reduce benchmark.* 19 files are staged from NFS to intermediate storage, 19 processes run in parallel, one per machine, each consuming one input file and producing one file that is tagged with *'collocation'*. These files are then consumed by a single workflow stage which writes a single file as output which is, then, staged out to NFS. The storage system stored staged-in files locally and prioritizes storing the files tagged with *'collocation'* on a single node, exposes data location, and the benchmark script uses this information to execute the reduce task on this node, avoiding the overhead of moving data.

*Scatter benchmark.* An input file is staged-in to the intermediate storage from NFS. The first stage of the workflow has one task that reads the input file and produces a scatter-file on intermediate storage. In the second stage, 19 processes run in parallel on different machines. Each process reads a disjoint region of the scatter-file and produces an output file. A tag specifies the block size to match the size of the application reading region (i.e., the region of the file that will be read by a process). Fine-grained block location information is exposed and enables scheduling the processes on the nodes that hold the block. Finally, at the stage-out phase, the 19 output files are copied to the back-end storage.

**Results.** Figure 5, Figure 6, Figure 7, and Figure 8 present the average benchmark runtime and standard deviation (over 20 runs) for five different intermediate storage systems setups. (1) NFS; (2) two setups for DSS - labeled 'DSS-RAM' or 'DSS-DISK' depending on whether the storage nodes are backed by RAM-disk or spinning disks; and (3) two setups for WOSS, labeled 'WOSS-RAM' or 'WOSS-DISK'). A sixth configuration is given for a local file system based on RAM-disk in the pipeline benchmark, representing the best possible performance.

Overall, a WOSS-* system exhibits the higher performance than the corresponding DSS-* system, which had better performance than NFS. This shows showing that the overheads brought by tagging, reading tags, and handling optimizations are paid-off by the performance improvements. Another advantage of WOSS is lower variance since it depends less on the network. Finally, as expected, RAM-disk-based configurations also perform faster and with less variability than their spinning disks counter-parts.

Locality in the pipeline scenario was the optimization that provided the best improvements. In this case, WOSS is 10x faster than NFS, 2x faster than DSS, and similar to *local* (the best possible scenario).

Staging and file creation for scatter benchmark take a significant amount of time (70-90%) of the benchmark time and, thus, for clarity of the presentation, the plot focuses only on the workflow stage that is affected by the optimization. Following the same trend of pipeline benchmark, scatter is 10.4x times faster than NFS and 2x faster than DSS. For reduce benchmark, DSS does not exhibit the same order of improvement over NFS. WOSS, however, is able to deliver almost 4x speedup compared to NFS.

The broadcast benchmark presents a more interesting case: Tagging for replication provides a finer tuning (number of replicas) for optimization than the other techniques that rely just on turning on/off (e.g., locality, and collocation). Figure 6 presents the performance for this benchmark when reaching the best performance (for 8 replicas). This result matches the expectation of the potential benefits of WOSS approach. For more replicas than optimal, the overhead of replication is higher than the gains.

In addition to using the workload presented in Figure 4 we also executed the benchmarks with data sizes scaled up (10x) and scaled down (1000x). The larger workload had results similar to the ones presented in this section. The smaller one did not show significant difference among the storage systems (less than 10%, in order of milliseconds) with DSS performing faster than WOSS in some cases since the overhead of adding tags and handling optimizations did not pay off for such smaller files.

### 4.2. Simple Real Applications: BLAST, modFTDock

This section evaluates WOSS for two relatively simple real-word workflow applications. We use:

- *modFTDock* [51] a protein docking application. The workflow has three stages with different data flow patterns (Figure 9): *dock* stage (broadcast pattern) verifies the similarity of molecules against a database, *merge* (reduce) summarizes the results for each molecule, and *score* (pipeline) produces a ranking for the molecules.

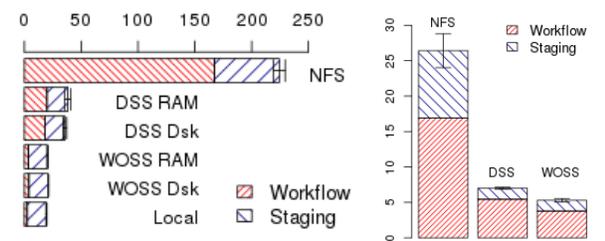

**Figure 5. Average time (in sec) for pipeline benchmark.**

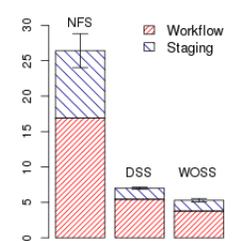

**Figure 6 - Average time (sec) for broadcast.**

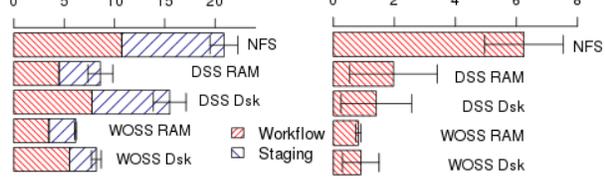

**Figure 7 - Average runtime for reduce benchmark**

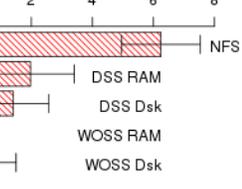

**Figure 8 - Average time (in sec) for stage 2 of scatter.**

- *BLAST* [52] a DNA search tool for finding similarities between DNA sequences. Each node receives a set of DNA sequences as input (a file for each node) and all nodes searches the same database file, i.e., BLAST has the broadcast pattern as the database has to be available on each application node for search. (Figure 12)

*Integration with workflow runtimes.* For *modFTDock* we use *Swift* to drive the workflow: *Swift* schedules each application stage, and tags the files according to the workflow pattern. As *modFTDock* combines the broadcast, reduce and pipeline pattern. *Swift* tags the database to be *replicated* (broadcast pattern), the output of every *dock* stages is *collocated* on a single storage node that will execute the *merge* stage (reduce). The merge output is tagged to be placed on local storage node in order to execute the *score* stage on the same machine (pipeline pattern). The labels on the arrows in Figure 9 indicate the hints used.

For *BLAST* we use shell scripting: the script that launches the *BLAST* experiment tags the database file to a specific replication level and the input file with the DNA sequences as '*local*'. The labels on the arrows in Figure 12 indicate the hints used.

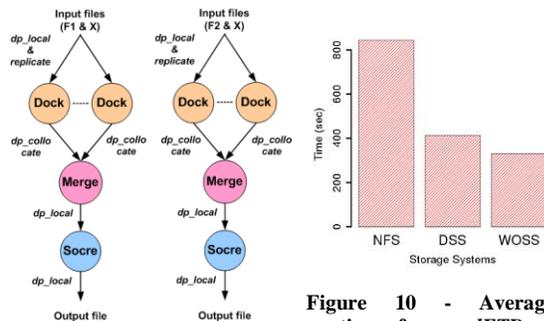

**Figure 9 - *modFTDoc* worflow**. Labels on arrows represent the tags used

**Figure 10 - Average runtime for *modFTDock* on cluster**. We run 9 pipelines in parralel using 18 nodes (average over 5 runs).

***modFTDock* experiments on cluster.** 9 dock streams progress in parallel and process the input files (100-200KB) and a database (100-200KB). The storage nodes are mounted on RAM-disks. Figure 10 presents the total execution time for the entire workflow including stage-in and stage-out times for DSS and WOSS. WOSS optimizations enable a faster execution: *modFTDock*/*Swift* is 20% faster when running on WOSS than on DSS, and more than 2x faster than when running on NFS.

***modFTDock* experiments on BG/P.** We ran *modFTDock* at larger scale on BG/P (Figure 11) to verify scalability and explore whether the performance gains are preserved when compared to a much more powerful backend storage (GPFS) available on this platform. On the one side, we notice a consistent 20-40% performance gain of DSS over GPFS. On the other side, we are not able to show positive results for WOSS: the application runtime is significantly longer than when using DSS. , We were able to attribute the performance loss to *Swift* runtime overheads introduced by *Swift* location aware scheduling, which is only used by WOSS, rather than to intrinsic WOSS overheads. We are currently exploring avenues to improve the scheduling algorithms.

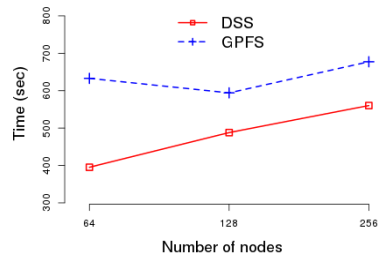

Figure 11. ***modFTDock* runtime on BG/P** while varying the number of nodes allocated to the application. The workload size increases proportionally with the resource pool.

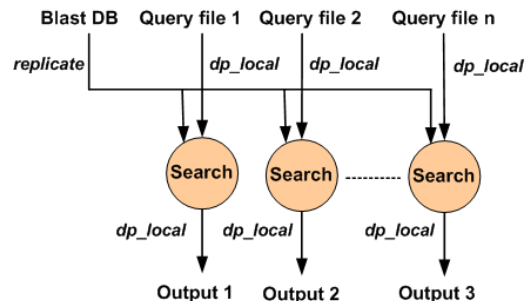

**Figure 12.** *BLAST* **workflow**. The *BLAST* database (1.8GB) is used by all nodes that search in parallel different queries. Labes on arrow represent the tags used to hint the data usage patters.

**Table 4. Average BLAST execution time** (in seconds) for NFS, DSS and various replication levels controlled in WOSS.

|  | NFS | DSS | WOSS (replication factor) | | | |
| --- | --- | --- | --- | --- | --- | --- |
|  |  |  | 2 | 4 | 8 | 16 |
| Stage-in | 49 | 17 | 19 | 29 | 36 | 55 |
| 90% workflow tasks | 264 | 185 | 164 | 155 | 151 | 145 |
| All tasks finished | 269 | 207 | 173 | 165 | 162 | 164 |
| Stage-out | 1 | 1.2 | 1.3 | 1.2 | 1.1 | 1.1 |
| **Total** | **320** | **226** | **193** | **191** | **200** | **221** |

**BLAST experiments on cluster.** 19 processes launch 38 DNA queries in the database independently and write results to backend storage. We report results from experiments that use a 1.7GB database. Output file sizes are 29 to 604KB. Table 4 presents the breakdown for the BLAST workflow runtime. Our approach does offer performance gains compared to NSF (up to 40% better performance) as well as compared to DSS (up to 15% better performance)

### 4.3. A Complex Workflow: Montage

The previous section demonstrated that the performance improvements highlighted by the synthetic benchmarks still hold under real simple workflow applications. This section evaluates the WOSS performance using a significantly more complex workflow application (Figure 13), Montage [18], with two goals in mind: First, we aim to evaluate the performance gains the cross layer optimizations approach bring to a real complex application, and secondly, we aim to understand in detail the overhead/performance gain balance brought by WOSS techniques (i.e., tagging, getting location information, location-aware scheduling) (next sub-section)

Montage [18] is an astronomy application that builds image mosaics from a number of independent images (e.g., smaller, or on different wavelength) captured by telescopes. The

Montage workflow is composed of 10 different processing stages with varying characteristics (Table 5). The workflow uses the reduce pattern in two stages and the pipeline patterns in 4 stages (as the labels in Figure 13 indicate).

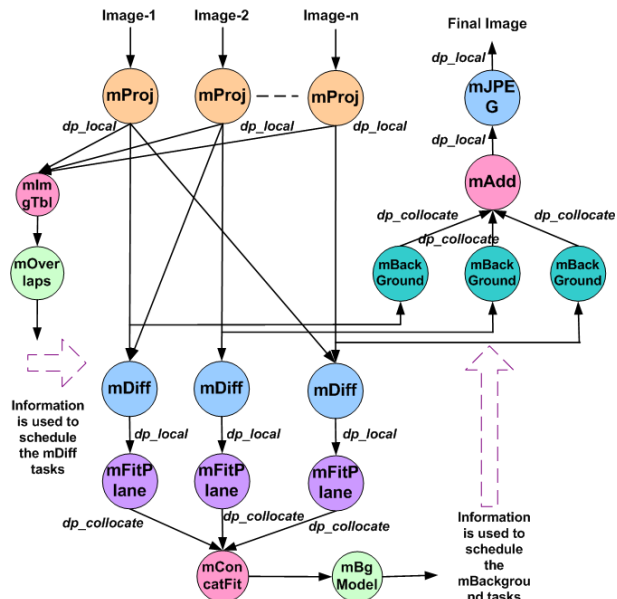

**Figure 13. Montage workflow**. The characteristics of each stage are described in Table 5. Labes on arrow represent the tags used to hint the data usage patters.

The I/O communication intensity between workflow stages is highly variable (presented in Table 5 for the workload we use). The workflow uses *pyFlow*. Overall the workflow generates over 650 files with sizes from 1KB to over 100MB and about 2GB of data are read/written from storage.

**Table 5: Characteristics of each stage for the Montage workflow**

| Stage | Data | #files | File size (per file) | Optimization |
|---|---|---|---|---|
| stageIn | 109 MB | 57 | 1.7 MB -2.1 MB | |
| mProject | 438 MB | 113 | 3.3 MB - 4.2 MB | Yes |
| mImgTbl | 17 KB | 1 | | |
| mOverlaps | 17 KB | 1 | | |
| mDiff | 148 MB | 285 | 100 KB - 3 MB | Yes |
| mFitPlane | 576 KB | 142 | 4.0 KB | Yes |
| mConcatFit | 16 KB | 1 | | |
| mBgModel | 2 KB | 1 | | |
| mBackground | 438 MB | 113 | 3.3 MB - 4.2 MB | Yes |
| mAdd | 330 MB | 2 | 165MB | Yes |
| mJPEG | 4.7 MB | 1 | 4.7 MB | Yes |
| stageOut | 170 MB | 2 | 170 MB | Yes |

Figure 14 shows the total execution time of the Montage workflow in five configurations: over NFS, and with DSS and WOSS deployed over the spinning disks. RAM-Disk experiments (not reported here) achieve similar results. The WOSS system achieves the highest performance when deployed on disk or RAM-disk. When deployed on disk WOSS achieves 30% performance gain compared to NFS. Further WOSS achieves up to 10% performance gain compared to DSS when deployed on disk or RAM-disk.

We also ran Montage on a larger cluster (50 nodes) on Grid5000. While WOSS achieves higher performance than NFS, it is comparable to DSS performance. We are still debugging this performance anomaly.

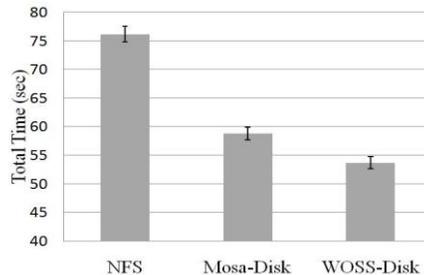

**Figure 14.** *Montage* **workflow execution time** (averages over four experiments). Note that, to better highlight the differences, y-axis does not start at zero.

### 4.4. Exploring WOSS Ovreheads/Gains

To enable cross layer optimization in WOSS a set operations are needed, including: forking a process to add a file extended attributes (labeled *fork* in Table 6), adding the extended attributes to files ('*tagging'*), getting the location for each file ('*get location'*), and performing location-aware scheduling ('*location-aware scheduling'*). All these steps add additional overhead and only after all three steps are done benefits can be reaped.

To guide future optimizations we run the same Montage workload as in the previous section yet configured to expose the overhead of each of these steps. For instance, to expose only the overhead of tagging, we tag the files, in all the benchmarks in Table 6 except WOSS, with a random tag that will add the overhead without triggering any optimization.

Table 6 shows the average execution time of the Montage workflow in these conditions. The results suggest that steps described above add significant overhead (up to 7%). A closer look reveals that the tagging operation is the main contributor to the overhead. The main reason behind this high overhead is twofold: first, every tagging operation incurs a roundtrip to the manager; second, the current manager implementation serializes all 'set-attribute' calls, this adds significant delay considering that Montage workflow produces and tags over 660 files in every run.

Our evaluation highlights that optimizing the 'set-attribute' operation (by caching and increasing the manager implementation parallelism) can bring significant additional (up to 7%) performance gains. We note that the use of fork was an implementation shortcut, we have evolved the code to use Pthon *xattrib* library in the meantime.

**Table 6. WOSS microbenchmark.**

| Experiment setup | Total time (s) |
|---|---|
| DSS | 66.2 |
| DSS + fork | 67.1 |
| DSS + fork + tagging | 69.5 |
| DSS + fork + tagging + get location | 70 |
| DSS + fork + tagging + get location + location-aware scheduling (on useless tags) | 70.7 |
| WOSS (all of the above with usefull tags) | 61.9 |

## 5. DISCUSSION AND SUMMARY

**Cross-layer optimizations** bypass a restricted, 'hourglass', interface between system layers. The key enabler is allowing information available at one layer to be visible at the other layer and drive optimizations. A classic example is the TCP/IP stack: in the original design, the transport layer makes the assumption that a lost packet is an indicator of congestion at the lower layers and backs-off. This assumption is violated in wireless environments and leads to degraded performance. To deal with this situation, a number of mechanisms have been designed to expose lower layers' state and channel capability [53, 54] such that the upper layer can infer the cause of packet loss and react appropriately.

Storage systems can be viewed through the same lens: the traditional (and after decades of use, convenient) POSIX file system API performs the role of the 'hourglass' neck: it enables transparent cooperation between applications and storage through a minimal interface. The POSIX interface, however, does not offer a mechanism to pass information between these layers. In the last two decades a number of systems proposed specialized APIs for passing applications hints to inform storage system optimizations. To date no widely used system adopts these approaches, as requiring changes to the standard API hinders adoption. This paper proposes using *standard extended* attributes in this role. We argue that this is a flexible, backward compatible, mechanism for communication between the storage and applications.

**Design guidelines.** Two design lessons relevant to storage system design can be borrowed from the design of the network stack: First, both applications and the storage should *consider metadata as hints rather than hard directives*. That is, depending on specific implementation and available system resources directives expressed through custom metadata might or might not be followed at all layers of the system. Second, to foster adoption, *adding support for cross layer optimizations should not (or minimally) impact the efficiency of applications or storage system not using them*. For example, if the top layer (an application) does not use the metadata offered by the lower layer, or decides not to pass hints, its performance should not be affected (otherwise these mechanisms are less likely to be adopted in practice as with some of the solutions that did not gain traction in the networking space [55])

We put forward two additional design guidelines: First, *the cross-layer communication and the optimizations enabled should not break the separation of concerns between layer*s. A key reason for layered designs is reducing system complexity by separating concerns at different layers. Therefore, it is necessary to devise mechanisms that limit the interference one layer may cause on others even though, as we argue, there are benefits in allowing information cross between layers. Second, the *distinction between mechanism and policy should be preserved*. The custom metadata offer a mechanism to pass information across layers. The various policies associated with the metadata should be kept independent from the tagging mechanism itself.

**Cross-layer optimizations in storage systems.** Apart from the usecase discussed in this paper a number of other uses of cross-layer optimizations based on custom metadata are possible. We briefly list them here:

*Cross-layer optimizations enabled by top-down (i.e., application to storage system) information passing.* Applications may convey hints to the storage system about their future usage patterns, reliability requirement, desired QoS, or versioning. *Future usage patterns:* there is a wealth of cross-layer optimizations that fit in this category apart from the ones we already explore. These include application-informed data prefetching, and data layout. *QoS requirements*: Different data items can have different, application-driven QoS requirements (e.g., access performance, availability or durability, and, possibly, security and privacy). A storage system that is aware of these requirements can optimize resource provisioning to meet individual items' QoS requirements rather than pessimistically provision for the most demanding QoS level. *Versioning*: Applications can use metadata to indicate a requirement to preserve past versions of a file. *Consistency requirements*: applications can use metadata to inform the storage system about their consistency requirements (e.g., by stating continuous consistency [56] requirements at the file level). Making the choice of the consistency requirements flexible allows the application to manage the tradeoffs between performance and consistency. *Energy optimization*: An energy optimized system may inform the storage system of the planned shutdown of subset of nodes in the system, the storage system can use this hint to inform the replication and consistency mechanism to avoid unnecessary replication and maintain consistency with replicas on shutdown nodes.

*Cross-layer optimizations enabled by bottom-up information passing.* In addition to exposing location information to enable effective scheduling decisions. Additional storage-level information (e.g., replication count, information about inconsistencies between replicas, properties and status of the storage device, caching status) could be useful when making application level-decisions as well (e.g., scheduling, data loss risk evaluation). For instance, exposing device specific performance characteristics can enable optimizing database operations [57], or optimizing the application I/O operations by matching the access pattern to the disk drive characteristics [13], or enable energy optimizations by exposing which nodes are contain less popular or well replicated blocks and can be shut down.

Similar to Tantisiroj et al. [38] and unlike many specialized storage system that advocate abandoning POSIX to enable extra functionality or enable higher performance (e.g. HDFS), we argue that storage systems can be specialized for certain applications while supporting POSIX. For instance, HDFS provides special API for getting the file location for location-aware scheduling, setting the replication level, or extracting file system statistics. All these operations can be expressed using our cross layer optimization approach to enable the same optimizations.

**Limitations.** The proposed approach and design have two main limitations. First, the proposed *per-file* cross layer optimization approach assumes that data of each file is stored separately from the other files, this limits the use of this approach in systems in which a single data block can be part of multiple files (e.g. content addressable storage, or copy-on-write storage system) as it is possible for separate files that share a block to have conflicting application hints. Second, our design allows extending the system and add optimization modules. This design decision is not accepted in secure storage system as it adds significant vulnerabilities.

**Summary.** This paper proposes using custom metadata as a bidirectional communication channel between applications and the storage system. We argue that this solution unlocks an incremental adoption path for cross layer optimizations in storage systems. We demonstrate this approach in context of workflow execution systems. Our workflow optimized storage system, exploits application hints to provide per-file optimized operations, and exposes data location to enable location-aware scheduling. The ssimple policies/hints we explore unlock sizeable performance benefits, suggesting that further work could yield bigger gains.

# 6. REFERENCES

[1] A. Parker-Wood et al., *Making Sense of File Systems Through Provenance and Metadata*, TR-UCSC- 12-01.
[2] S. Jones, et al. *Easing the Burdens of HPC File Management*. in *Parallel Data Storage Workshop*. 2011.
[3] S. Ames et al., *Richer file system metadata using links and attributes*. MSST 2005.
[4] S. Ames, et al. *LiFS: An attribute-rich file system for storage class memories* MSST. 2006.
[5] J. Koren, et al. *Searching and Navigating Petabyte Scale File Systems Based on Facets*. ACM PDSW 2007
[6] A. W. Leung, et al. *Spyglass: Fast, Scalable Metadata Search for Large-Scale Storage Systems*. in *FAST*. 2009.
[7] K.-K. Muniswamy-Reddy, et al. *Provenance-Aware Storage Systems*. in *USENIX ATC*. 2006.
[8] A. C. Arpaci-Dusseau, et al. *Transforming Policies into Mechanisms with Infokernel*. in *SOSP*. 2003.
[9] G. Fedak et al., *BitDew: a programmable environment for large-scale data management*. *SC*. 2008.
[10] H. Song et al., *Cost-intelligent data layout optimization for parallel filesystems.* Cluster Computing, 2012.
[11] A. C. Arpaci-Dusseau, et al., *Semantically-smart disk systems: past, present, and future.* SIGMETRICS Performance Evaluation Review, 2006.
[12] G. Alonso et al., *SwissBox: An Architecture for Data Processing Appliances*. CIDR 2011.
[13] J. Schindler, et al.,. *Track-aligned Extents: Matching Access Patterns to Disk Drive Characteristics*. FAST'02.
[14] M. Wilde, et al., *Swift: A language for distributed parallel scripting.* Parallel Computing, 2011.
[15] E. Santos-Neto et al., *Enabling Cross-Layer Optimizations in Storage Systems with Custom Metadata*. in *HPDC - Hot Topics Track*. 2008.
[16] E. Vairavanathan et al., *A Workflow-Aware Storage System: An Opportunity Study*. *CCGrid*. 2012.
[17] *modFTDock*. 2011; http://www.mybiosoftware.com/3d-molecular-model/922.
[18] A. Laity et. al.,. *Montage: An Astronomical Image Mosaic Service for the NVO*, *ADASS*. 2004.
[20] I. Raicu, et al., *Many-Task Computing for Grids and Supercomputers*, MTAGS. 2008.
[21] Z. Zhang, et al. *Design of Collective I/O for Loosely-coupled Petascale Programming*. MTAGS. 2008.
[22] J. Bent, et al. *Explicit Control in a Batch-Aware Distributed File System*. NSDI 2004..
[23] M. Li, et al. *Functional Partitioning to Optimize End-to-End Performance on Many-core Architectures*. *SC* 2010
[24] H. M. Monti, et. al., *Reconciling Scratch Space Consumption, Exposure, and Volatility to Achieve Timely Staging of Job Input Data*. IPDPS. 2010.
[25] S. Boboila, et al. *Active Flash: Out-of-core Data Analytics on Flash Storage*. in *MSST*. 2012.
[26] J. Wozniak et al., *Case studies in storage access by loosely coupled petascale applications*, PDSW 2009.
[27] D. S. Katz, et al., *Many-Task Computing and Blue Waters*, Technical Report CI-TR-13-0911.
[28] T. Shibata, et al., *File-access patterns of data-intensive workflow applications and their implications to distributed filesystems*, HPDC 2010.
[29] S. Bharathi, et al., *Characterization of Scientific Workflows*, WWSLSS 2008.
[30] U. Yildiz, et al., *Towards scientific workflow patterns*, in *WWSLSS* 2009.
[31] S. Ghemawat, H. Gobioff, and S.-T. Leung. *The Google File System*. in *SOSP*. 2003.
[32] K. Shvachko, H. Kuang, S. Radia, and R. Chansler. *The Hadoop Distributed File System*. in *MSST* 2010.
[33] K. Gupta et al., *GPFS-SNC: An enterprise storage framework for virtual-machine clouds* IBM Journal of Research and Development 2011.
[34] M. Rosenblum et al, *The Design and Implementation of a Log-Structured File System*. in *ACM TCS* 1992.
[35] MetaFS, http://metafs.sourceforge.net/. 2012.
[36] E. Adar et al.,. *Haystack: Per-User Information Environments*. in *ICDM* 1999.
[37] C. Maltzahn et al. *Graffiti: A Framework for Testing Collaborative Distributed Metadata*. *Informatics*. 2007.
[38] W. Tantisiriroj, et al. *On the Duality of data-intensive filesystem design: reconciling HDFS and PVFS*, *SC*'11.
[39] M. P. Mesnier and J. B. Akers, *Differentiated storage services.* SIGOPS OS Review, 2011. **45**(1).
[40] Z. Chen et al. *Empirical evaluation of multi-level buffer cache collaboration for storage systems*. in *SIGMETRICS*. 2005.
[41] R. H. Patterson, et al. *Informed prefetching and caching*. in *SOSP*. 1995.
[42] K. Fujimoto et al. *Power-aware Proactive Storage-tiering Management for High-speed Tiered-storage Systems*. in *WSIT* 2010.
[43] N. Mandagere et al. *GreenStor: Application-Aided Energy-Efficient Storage*. in *MSST*. 2007.
[44] M. Abd-El-Malek et al. *Ursa minor: versatile cluster-based storage*. in *FAST* 2005.
[45] *SNIA XAM Initiative*. [cited 2010; http://www.snia.org/forums/xam/.
[46] P. H. Carns, et al., *PVFS: A Parallel File System for Linux Clusters*. *4th Annual Linux Showcase, 2000*.
[47] F. Schmuck and R. Haskin. *GPFS: A Shared-Disk File System for Large Computing Clusters*. in *FAST*. 2002.
[48] D. Rochberg et al, *Prefetching Over a Network: Early Experience With CTIP*. SIGMETRICS PER. 1997.
[49] *FUSE,* [cited 2011; http://fuse.sourceforge.net/.
[50] BG/P team, *Overview of the IBM Blue Gene/P Project.* IBM Journal of Research and Development, 2008. **52**
[52] S. F. Altschul, et al., *Basic Local Alignment Search Tool.* Molecular Biology, 1990. **215**: p. 403–410.
[53] A. Gurtov et al., *Modeling wireless links for transport protocols.* ACM SIGCOMM CCR 2004
[54] A. Gurtov et al., *Lifetime packet discard for efficient transport over cellular links.* ACM SIGMOBILE 2003.
[55] R. Fonseca, et al., *IP Options are not an option*, in *Technical Report No. UCB/EECS-2005-24*. 2005.
[56] A. S. Tanenbaum and M. V. Steen, *Distributed Systems: Principles and Paradigms*. 2 ed. 2006: Prentice Hall.
[57] J. Schindler et al., *Lachesis: Robust Database Storage Management Based on Device-specific Performance Characteristics*. VLDB 2003.